\documentclass[lettersize,journal]{IEEEtran}
\usepackage{amsmath,amsfonts}
\usepackage{array}
\usepackage{textcomp}
\usepackage{stfloats}
\usepackage{url}
\usepackage{verbatim}
\usepackage{graphicx}
\usepackage[
maxnames=10,
giveninits=true
]{biblatex}

\usepackage{color}

\usepackage{multirow}
\usepackage[normalem]{ulem}
\useunder{\uline}{\ul}{}
\usepackage{subfigure}
\newtheorem{myDef}{Definition}
\usepackage[ruled]{algorithm2e}

\begin{document}
\title{Predictive and Contrastive: Dual-Auxiliary Learning for Recommendation}

\author{Yinghui~Tao, Min~Gao, Junliang~Yu, Zongwei~Wang, Qingyu~Xiong, Xu~Wang
\thanks{This work was supported by National Key R\&D Program of China (2018YFB1403602), the National Natural Science Foundation of China (62176028), and the Overseas Returnees Innovation and Entrepreneurship Support Program of Chongqing (cx2020097). (\textit{Corresponding author: Min Gao.})}

\thanks{Yinghui Tao, Min Gao, Zongwei Wang, and Qingyu Xiong are with the Key Laboratory of Dependable Service Computing in Cyber Physical Society (Chongqing University), Ministry of Education and School of Big Data and Software Engineering, Chongqing 401331, China (e-mail: taoyinghui@cqu.edu.cn; gaomin@cqu.edu.cn; zongwei@cqu.edu.cn; cquxqy@163.com).}

\thanks{Junliang Yu is with the School of Information Technology and Electrical Engineering, The University of Queensland, Queensland 4072, Australia (e-mail: jl.yu@uq.edu.au).}


\thanks{Xu Wang is with the College of Mechanical and Vehicle Engineering, Chongqing University, Chongqing 400044, China (e-mail: wx921@163.com).}
}



\maketitle

\begin{abstract}
Self-supervised learning (SSL) recently has achieved outstanding success on recommendation. By setting up an auxiliary task (either predictive or contrastive), SSL can discover supervisory signals from the raw data without human annotation, which greatly mitigates the problem of sparse user-item interactions. However, most SSL-based recommendation models rely on general-purpose auxiliary tasks, e.g., maximizing correspondence between node representations learned from the original and perturbed interaction graphs, which are explicitly irrelevant to the recommendation task. Accordingly, the rich semantics reflected by social relationships and item categories, which lie in the recommendation data-based heterogeneous graphs, are not fully exploited. To explore recommendation-specific auxiliary tasks, we first quantitatively analyze the heterogeneous interaction data and find a strong positive correlation between the interactions and the number of user-item paths induced by meta-paths. Based on the finding, we design two auxiliary tasks that are tightly coupled with the target task (one is predictive and the other one is contrastive) towards connecting recommendation with the self-supervision signals hiding in the positive correlation. Finally, a model-agnostic DUal-Auxiliary Learning (DUAL) framework which unifies the SSL and recommendation tasks is developed. The extensive experiments conducted on three real-world datasets demonstrate that DUAL can significantly improve recommendation, reaching the state-of-the-art performance.
\end{abstract}


\section{Introduction}
\IEEEPARstart{M}{ost} modern recommender system models are based on deep neural architectures, which require a large volume of training data to take full advantage of their capacity. However, since users can only interact with a fraction number of provided items, the observed user behavioral data is extremely sparse, making deep neural recommendation models struggle to learn high-quality representations \cite{SGL}. Recently, self-supervised learning (SSL) becomes a latest trend in multiple fields. Due to its effectiveness to discover supervisory signals from the raw data without human annotation, it is inherently a sliver bullet to the data sparsity issue, and has gained attention from the recommendation community \cite{XieY}.

Some early research effort on SSL has transplanted the successful ideas from other fields to recommendation. For example, Zhou \emph{et al.} \cite{S3-rec} introduce SSL into the sequential recommendation to learn the intrinsic correlation of data through maximizing the mutual information between different types of data. Wu \emph{et al.} \cite{SGL} improved the robustness and effect of the recommendation model by generating different views for nodes and then self-discriminating them. The common thought of these works is to set up an auxiliary task for mining self-supervision signals. However, despite their decent performance, since these auxiliary tasks are derived from other scenarios such as language modeling and graph learning, they cannot seamlessly be generalized to the recommendation scenario. 

\begin{figure}[htbp]
\centering
\includegraphics[width=0.45\textwidth]{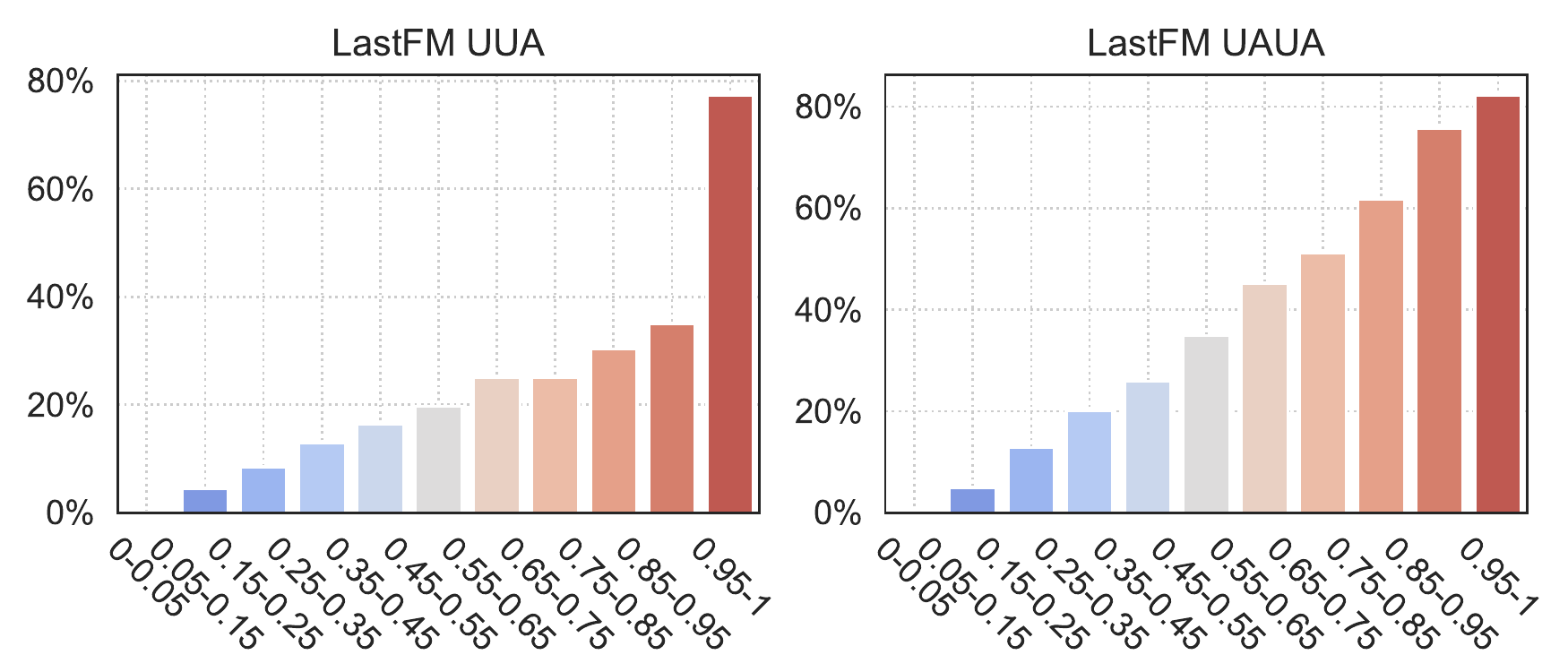} 
\caption{The positive correlation between user-item interactions and the induced path numbers via meta-paths on Last-FM. The horizontal axis represents the link-score derived from the number of path instances between the head user and the tail artist; the vertical axis represents the possibilities with which the head user consumed the tail item in the paths.}    
\label{lastfm}       
\end{figure}

Most real-world scenarios contain multiple node types that need to be converted into heterogeneous graphs \cite{HeterogeneousGraph}. However, most research on self-supervised learning for recommender systems focuses on homogeneous graphs with a single type of node and edge. Accordingly, the rich semantics lying in heterogeneous graphs constructed from recommendation data, such as users' social relationships and item categories, are not fully exploited in these tasks. Meta-paths are introduced as a representative instantiation of heterogeneous graphs to learn multi-hop semantic embeddings. Consequently, we capture the semantic information by deriving various meta-paths starting with a head node (user), passing through several objects, and then terminating at a tail node (item). To design recommendation-specific auxiliary tasks, we first quantitatively analyze the interaction data of the public music recommendation dataset Last-FM and find a strong positive correlation between the interactions and the numbers of user-item paths induced by meta-paths. Fig.\ref{lastfm} shows the analysis of meta-paths User-User-Artist (UUA) and User-Artist-User-Artist (UAUA). Consistent with intuition, the higher the number of path instances between a user and an artist, the higher the likelihood of an interaction between them. Based on this finding, we innovatively design a path-regression task to predict the link-score, which was extracted from the number of meta-path instances. More precisely, we construct the commuting matrix \cite{Meta-pathCommutingMatrix} of a specific meta-path instances, normalize it to obtain the link-score. Subsequently, the link-score is used as the self-supervision signal for a path-regression task, fully integrating semantic information and improving recommendation.


We obtain two discriminative representations for each node through the path-regression and recommendation tasks. For the sake of refining the indispensable information existing in the user-item bipartite graph, we employ a contrastive objective that distinguishes the representations derived from the path-regression and recommendation tasks to make representations more accurate. The path-regression task belongs to the predictive task of self-supervised learning \cite{XieY}. By combining two types of self-supervised tasks (predictive and contrastive tasks), our model self-generates informative supervision and handles the data-label relationships on the one hand, and deals with the inter-data information (data-data pairs) on the other hand.

To unify the recommendation task and the two auxiliary tasks, we jointly optimize their objectives. However, the competition and conflict between different tasks may lead to performance deterioration, namely negative transfer \cite{transfer}. In light of this, we propose a self-supervised \underline{Du}al-\underline{A}uxiliary \underline{L}earning framework (DUAL) in this paper. The framework adopts the Customized Gate Control \cite{HongyanTang} as the bottom architecture, which explicitly separates shared and task-specific components to avoid inherent conflicts and negative transfer, so that the recommendation performance can be significantly improved. The major contributions of this paper are summarized as follows:
\begin{itemize}
\item We analyze the interaction data and find a strong positive correlation between the interactions and the numbers of user-item paths induced by meta-paths, which provides reliable self-supervision signals for model training. 
\item We propose a model-agnostic self-supervised dual-auxiliary learning framework. It is with good extensibility, and can empirically adapt to different kinds of graph neural networks. 
\item Extensive experiments demonstrate the superiority of DUAL by comparing it with the state-of-the-art supervised and self-supervised models. Furthermore, this work experimentally verifies that jointly applying path-regression predictive and contrastive tasks can fully integrate semantic information and significantly improve model performance.
\end{itemize}

\section{Related Work}

\subsection{Self-Supervised Learning in Recommender Systems}
Most work in recommender systems has focused on supervised or semi-supervised learning settings, which require adequate user-item interaction data for model training. However, interaction data is often limited, expensive, and inaccessible. These works are challenging to adapt to real-world scenarios owing to their severe dependence on quantity interactions.

In such cases, self-supervised learning \cite{TianshengYao, HHGR} emerges as the times require. It aims to train a model on an auxiliary task where signals are automatically obtained from available data, withdrawing the need for excessively user-item interactions. Based on how the auxiliary tasks are designed, Xie \emph{et al.} \cite{XieY} divide the SSL methods into two categories, namely contrastive models and predictive models. The major difference between the two categories is that contrastive models require data-data pairs for training, while predictive models require data-label pairs, where the labels are self-generated from the data. Contrastive tasks aim to construct data-data pairs to refine the information between the data. Predictive tasks seek labels beneficial to learning node preferences from the original data and then learn the relationship between data and labels.

Inspired by the success of SSL, some recent work \cite{TianshengYao, JianxinMa, S3-rec} has applied it to recommender systems. Yao \emph{et al.} \cite{TianshengYao} add a regularization term to a large-scale item recommendation model by perturbing item embeddings. Ma \emph{et al.} \cite{JianxinMa} supplement future sequences from the perspective of long-term item sequence prediction to solve the problem of sequence recommendation. Lee \emph{et al.} \cite{BUIR} used only positive sample pairs as input to the model, making the embeddings of users and items with interactive behaviors closer to each other. In addition, the training data is supplemented utilizing random data augmentation. Zhang \emph{et al.} \cite{HHGR} thoroughly consider the complex relationship between users and groups, using two random data enhancements with different granularities for group recommendation scenarios. Xia \emph{et al.} \cite{xia2021self} proposed a self-supervised hypergraph convolution method for the session-based recommendation, which maximizes the mutual information between the line graph and transformed hypergraph. Yu \cite{yu2021self} \emph{et al.} use hypergraphs to model social relations and purify user embeddings by maximizing mutual information at the user level. Besides, some works attempt to utilize side-information of items, e.g., item attributes, in pretraining \cite{S3-rec}. Unlike previous works that merely focus on general-purpose contrastive tasks, we designed two auxiliary tasks closely related to the recommendation to fully integrate rich semantic information reflected by social relationships and item categories during model training.

\subsection{Recommender Systems Based on Meta-path}

Meta-path \cite{Meta-pathCommutingMatrix} is a structure to capture the semantics and has been widely applied in recommender systems. Zheng \emph{et al.} \cite{JingZheng} utilize meta-path-based similarity to normalize node embeddings and force more similar node pairs closer. Shi \emph{et al.} \cite{ChuanShi} first learn node embeddings along each path and then fuse them. Hwang \emph{et al.} \cite{Hwang} propose to assist recommendation by predicting whether user-item pairs can be connected through specific meta-paths. However, these models lack explicit feedback to guide the training of models. Implicit feedback is binary and naturally noisy. The interactions between users and an item do not mean users like the item, which may be due to a mistake click. On the contrary, users might not click on an item because they did not notice it, which does not mean they will alienate it. Compared with implicit feedback, explicit feedback reflects users’ preferences for items in a more fine-grained manner. 

Consistent with previous work, we likewise utilize meta-paths to support learning user preferences. Nonetheless, unlike previous works, our proposed model automatically provides valuable explicit feedback for model training.

\subsection{Recommender Systems Based on Multi-Task Learning}

In this work, we apply the joint training of target and auxiliary tasks to optimize our model. Multi-task learning (MTL) \cite{ZhaoZ} aims to guide model training by designing different objectives, i.e., training multiple tasks that affect each other simultaneously in a unified model \cite{HongyanTang}. MTL is widely used in recommendation scenarios to exploit various user behaviors better and achieve substantial improvement. Multiple related tasks are trained together, and they can share the information they learn during the training process.

Some studies \cite{li2021random, li2020joint} jointly model the internal relationship between social behavior and purchasing behavior to promote the two behavior prediction tasks simultaneously. In addition, some recommender systems have applied MTL models with efficient shared learning mechanisms. One way is to divide the bottom layer of the model into multiple independent modules \cite{HongyanTang}. Then, the top-level module responsible for predicting specific tasks fuses the output of the bottom-level modules. Ma \emph{et al.} \cite{JiaqiMa} introduce Mixture-of-Experts to MTL and the Multi-gate Mixture-of-Experts (MMOE) model to selectively fuse the outputs of the bottom-level modules by adding a gating mechanism for each task. Zhao \emph{et al.} \cite{ZhaoZ} analyze the effectiveness of various soft parameter sharing methods for optimizing multi-task learning. Unlike MMOE Multi-gate Mixture-of-Experts (MMoE) treat all bottom-level modules equally and draw information from them for each task, Tang \emph{et al.} \cite{HongyanTang} propose the CGC model that divides the bottom-level modules. Some modules are public, and others only affect a specific task. CGC is a state-of-the-art MTL method. It improves shared learning efficiency and addresses negative transfer further \cite{transfer}. Accordingly, we take it as the basic framework of our model.

\section{Motivation}

\begin{figure}[htbp]
\centering
\subfigure[An illustrative example of user-item bipartite graph.]
{  
\includegraphics[scale=0.35]{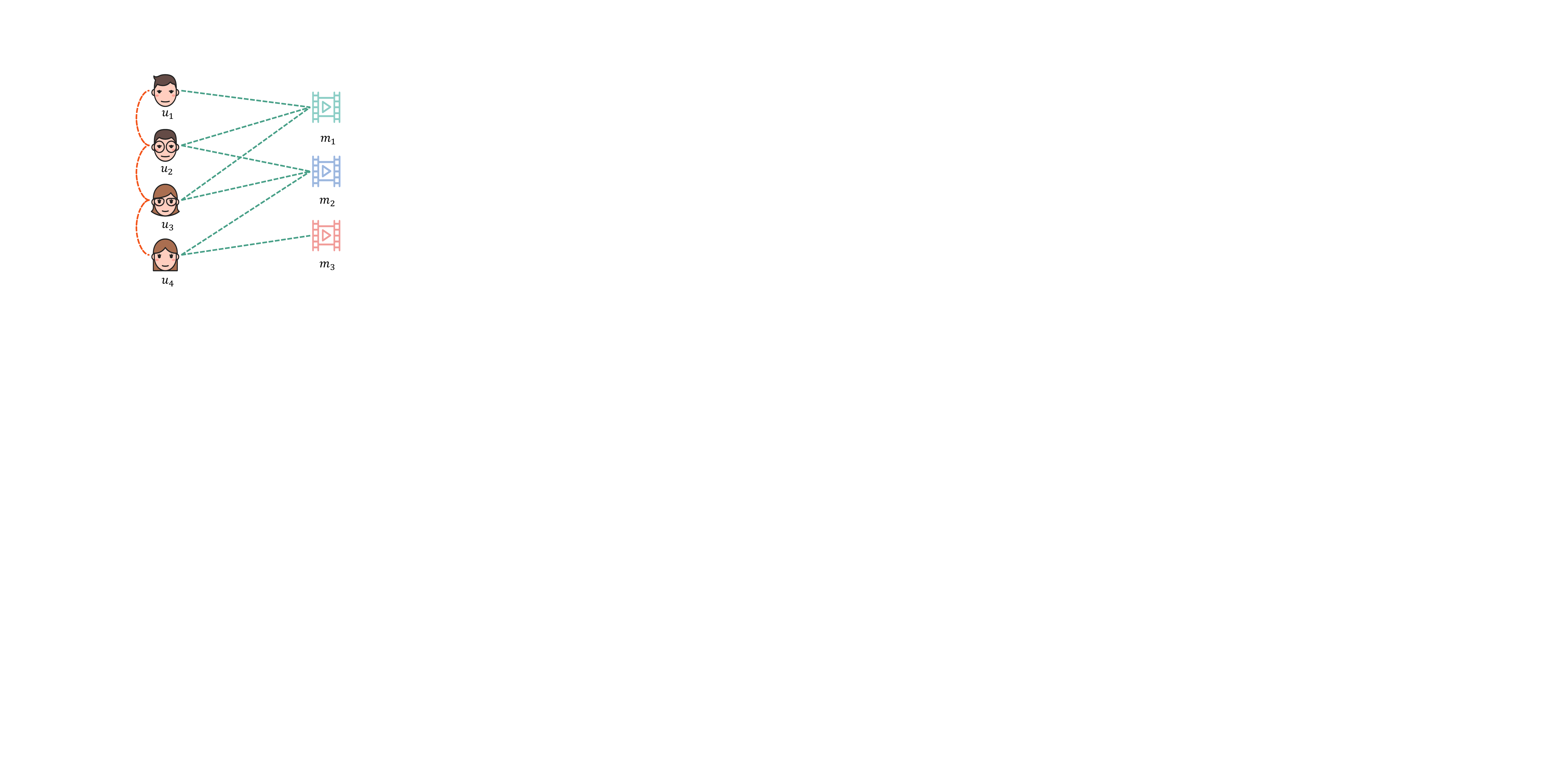}  
}
\subfigure[An illustrative example of the motivation.]
{  
\includegraphics[scale=0.35]{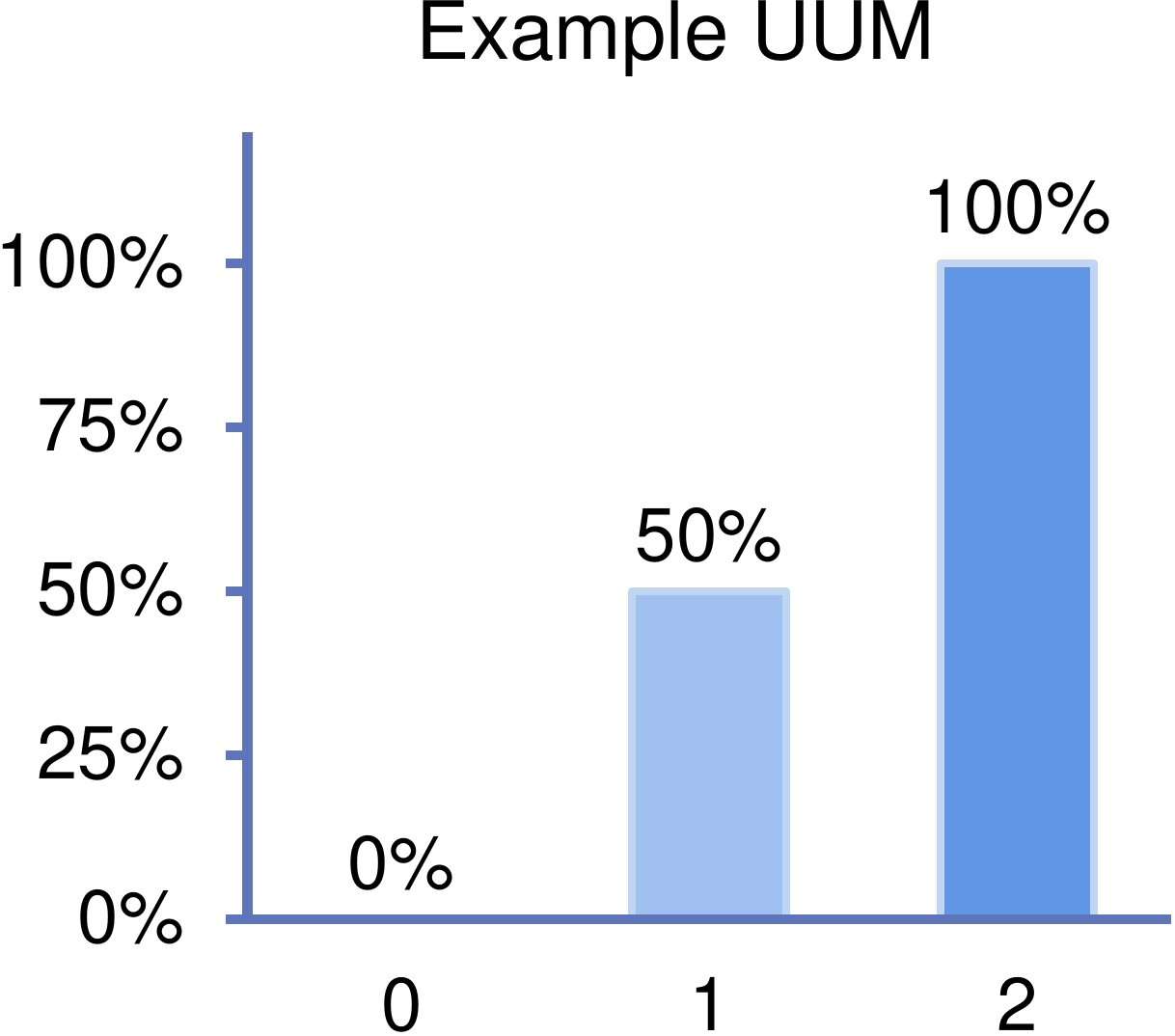}
}
\caption{The positive correlation between user-item interactions and the induced path numbers via meta-path UUM.}    
\label{example}       
\end{figure}

Because heterogeneous graphs contain more comprehensive information, such as users’ social relationships and item categories, they are widely applied in various data mining scenarios \cite{zhou2020deep, yang2018enriching, li2021random, li2020joint}. As a typical instantiation of heterogeneous graphs, meta-path \cite{Meta-pathCommutingMatrix} is widely applied to capture the semantics reflected from complex information. Taking the movie data in Fig.\ref{example}(a) as an example, the meta-path User-User-Movie (UUM) indicates that users’ preferences for movies be affected by their social connections, while User-Movie-User-Movie (UMUM) indicates that users explore new movies through other users who have seen the same film. Different semantic information in the recommended scenario can be extracted through the difference of node types passed by the meta-path.

The recommender system model based on meta-paths has already made many attempts \cite{JingZheng, ChuanShi, personality}. The work of these researchers shows that meta-paths help to learn users’ preferences for items. From this conclusion, we suspect that embeddings that can accurately predict the number of path instances are also helpful for the model to judge users' preferences.

We perform an analysis on the public music recommendation dataset Last-FM based on this idea. The number of path instances following a specific meta-path of user-item can be obtained by extracting the commuting matrix for the meta-path. Then we normalized the connection number in the matrix to a link-score of [0-1] to eliminate the differences among meta-paths. Fig.\ref{lastfm} shows the result: the more connections between user and artist following meta-path User-User-Artist (UUA) and User-Artist-User-Artist (UAUA), the higher the possibility that the user will interact with the artist. In addition, similar phenomena can also be observed in other datasets. The experimental results confirmed the conjecture mentioned above.

To facilitate understanding, we take the movie data constructed artificially in Fig.\ref{example} as an example:
\textit{
\begin{enumerate}[]
\item $u_{2}$ has \textbf{2} friends who watched $m_{1}$ and $u_{2}$ \textbf{watched} $m_{1}${\rm;}
\item $u_{3}$ has \textbf{2} friends who watched $m_{2}$ and $u_{3}$ \textbf{watched} $m_{2}${\rm;}
\item $u_{1}$ has \textbf{1} friends who watched $m_{1}$ and $u_{1}$ \textbf{watched} $m_{1}${\rm;}
\item $u_{3}$ has \textbf{1} friends who watched $m_{3}$ but $u_{3}$ \textbf{did not watch} $m_{3}${\rm;}
\item $u_{2}$ has \textbf{0} friends who watched $m_{3}$ but $u_{2}$ \textbf{did not watch} $m_{3}${\rm.}
\end{enumerate}
}

As shown in Fig.\ref{example}(b), the horizontal and vertical axes represent the induced path numbers and the possibilities of interactions between the head user and the tail item, respectively. It can be found that the number of path instances between user and movie following the UUM meta-path is related to whether the user to watch the movie, specifically:

\begin{itemize}
\item \textit{Users with \textbf{two} friends who watched the same movie are \textbf{100\%} likely to watch the movie{\rm;}}
\item \textit{Users with \textbf{one} friend who watched the same movie are \textbf{50\%} likely to watch the movie{\rm;}}
\item \textit{Users with \textbf{no} friends who watched the same movie are \textbf{0\%} likely to watch the movie{\rm.}}
\end{itemize}

We design two self-supervised auxiliary tasks for recommendation based on this phenomenon.


\section{Preliminary}
 The data extracted in real-world is usually not of a single type but contains many different types of nodes. The relationships between the nodes are also diverse. It is necessary to convert these data into heterogeneous graphs and use tools such as meta-paths to model complex relationships between nodes. This section first introduces relevant concepts in this work and then presents the problem statement.
\subsection{Relevant Concepts}

\begin{myDef}
\textbf{Heterogeneous Graph} \cite{HeterogeneousGraph}. 
Let $\mathcal{A}$ be the set of node types, and $\mathcal{T}$ be the set of edges types. The node and edge sets are denoted as $\mathcal{V}$ and $\mathcal{E}$, respectively. Then the heterogeneous graph containing $\phi: \mathcal{V} \rightarrow \mathcal{A}$ and $\psi: \mathcal{E} \rightarrow \mathcal{T}$ mapping functions is denoted as $\mathcal{G}=(\mathcal{V}, \mathcal{E})$, where $|\mathcal{A}|+|\mathcal{T}|>2$.
\end{myDef}

\begin{myDef}
\textbf{Meta-path} \cite{Meta-pathCommutingMatrix}.
A meta-path $\Phi$ is defined on the heterogeneous graph $\mathcal{G}=(\mathcal{V}, \mathcal{E})$, denoted as a path pattern in the form of $A_{1} A_{2} \cdots A_{l+1}$, which describes the composite semantic between the head node $A_{1}$ and the tail node $A_{l+1}$.
\end{myDef}

\begin{myDef}
\textbf{Commuting Matrix} \cite{Meta-pathCommutingMatrix}. 
A Commuting Matrix $\boldsymbol{C}$ is defined on the heterogeneous graph $\mathcal{G}=(\mathcal{V}, \mathcal{E})$ and the meta-path $\Phi=A_{1} A_{2} \cdots A_{l+1}$. The adjacency matrix between the node of type $A_{i}$ and the node of type $A_{j}$ is denoted as $\boldsymbol{W}^{i j} \in \{0,1\}^{\left|\mathcal{V}_{A_{i}}\right| \times\left|\mathcal{V}_{A_{j}}\right|}$, where $\mathcal{V}_{A_{i}}$ denote as the set of nodes of type $A_{i}$. $W_{m n}^{i j}=1$ represents there is a connection between node $m$ of node type $A_{i}$ and node $n$ of node type $A_{j}$, otherwise $W_{m n}^{i j}=0$. The commuting matrix $\boldsymbol{C}$ of meta-path $\Phi=A_{1} A_{2} \cdots A_{l+1}$ is defined as $\boldsymbol{C}^{\Phi}=\boldsymbol{W}^{12} \boldsymbol{W}^{23} \cdots \boldsymbol{W}^{l(l+1)}$.

\end{myDef}

\begin{myDef}
\textbf{Link-score Matrix}. A Link-score Matrix $\boldsymbol{M}$ is defined on the heterogeneous graph $\mathcal{G}=(\mathcal{V}, \mathcal{E})$ and the meta-path $\Phi=A_{1} A_{2} \cdots A_{l+1}$. Given the commuting matrix $\boldsymbol{C}^{\Phi} \in \mathbb{R}^{\left|\mathcal{V}_{user}\right| \times\left|\mathcal{V}_{item}\right|}$ of meta-path $\Phi$, the link-score matrix $\boldsymbol{M}^{\Phi} \in \mathbb{R}^{\left|\mathcal{V}_{user}\right| \times\left|\mathcal{V}_{item}\right|}$ is defined as a matrix obtained by normalizing commuting matrix $\boldsymbol{C}^{\Phi}$. 
\end{myDef}

The paths between two nodes are infinite, but the paths that follow a particular meta-path are finite. Therefore, for arbitrary user-item pairs, the number of path instances induced by meta-path $\Phi$ can be obtained by computing the commuting matrix $\boldsymbol{C}^{\Phi}$.

The data in recommender systems are often extremely sparse. Therefore, we can use sparse matrixes to store and operate when we calculate commuting matrix, i.e., only non-zero elements in the matrix are considered. In addition, the link-score matrix needs to be calculated only once and does not be updated iteratively. Therefore, the time consumption of calculating the link-score is acceptable.

As for the auxiliary task, we use the link-score matrix $\boldsymbol{M}$ as explicit supervisory signals of auxiliary tasks during model training. Taking the commuting matrix $\boldsymbol{C}$ and link-score matrix $M$ in Fig.\ref{example} as an example, $C^{\Phi}_{i, j}$ represents the number of path instances between $i$-th user and $j$-th item following the meta-path $\Phi$. Clearly, $C_{2, 1}^{UUM}=2$ represents that there are 2 path instances (i.e. $u_{2}{\longrightarrow}u_{1}{\longrightarrow}m_{1}$ , $u_{2}{\longrightarrow}u_{3}{\longrightarrow}m_{1}$) between $u_{2}$ and $m_{1}$ following the meta-path $UUM$. Similarly, $C_{2, 2}^{UUM}=1$, $C_{2, 3}^{UUM}=0$. To eliminate the magnitude difference in the number of path instances between users, we normalize the commuting matrix according to rows (users) to obtain the link-score, i.e. $M_{2, 1}^{UUM}=1$, $M_{2, 2}^{UUM}=0.5$, and $M_{2, 3}^{UUM}=0$. As a result, although active users have numerous meta-paths, it does not lower the link-score of inactive users.

\subsection{Problem Statement}

Consider a recommender system with a user-set $\mathcal{U} = \left \{ u_{1},\ldots,u_{m}\right \}$ and an item-set $\mathcal{I} = \left \{ v_{1},\ldots,v_{n}\right \}$. A sparse interaction matrix $\boldsymbol{R} \in \mathbb{R}^{m \times n}$ is used to denote all users’ implicit feedback on all items, where each element $r_{uv}$ = 1 indicates that user $u$ has observed or interacted with item $v$ before, and $r_{uv}$ = 0 otherwise. The goal of the recommendation model is to to provide users with a Top-K item recommendation list based on historical interactions. 

Graph-based methods usually denote the user-item bipartite graph as $\mathcal{G}_{UI} = \left \{ (u, v)|u \in \mathcal{U}, v \in \mathcal{I} \right \}$, which is constructed from user-item history interaction matrix $\boldsymbol{R}$. A nonzero $r_{uv}$ matches an edge between user $u$ and item $v$ on $\mathcal{G}_{UI}$. Besides, we denote the user-item bipartite graph as $\mathcal{G}_{\Phi} = \left \{ (u, v)|M_{u,v} > 0 \right \}$, which is constructed from the link-score matrix $\boldsymbol{M}$ of meta-path $\Phi$. The two embedding matrices initialized for all nodes, generated by parameterizing each ID with a random initial vector, are denoted as shared embedding $\mathbf{E}_{S} \in \mathbb{R}^{(m+n) \times d}$ and task-specific embedding $\mathbf{E}_{T} \in \mathbb{R}^{(m+n) \times d}$, respectively. $d$ is the embedding dimension. 

Given a target user, our goal is to generate a ranked list over the item-set $\mathcal{I}$ for the given user.




\begin{figure*}[htbp]
\centerline{\includegraphics[scale=0.38]{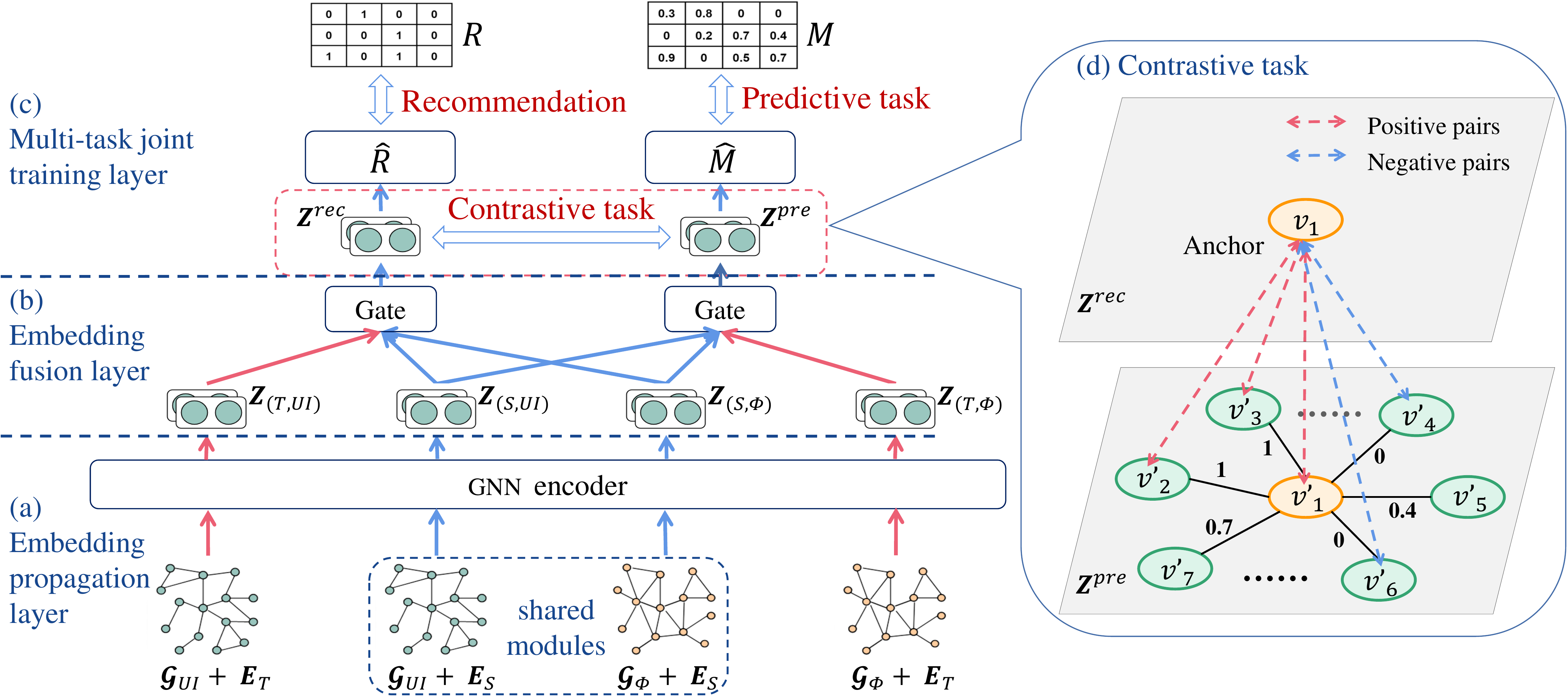}}
\caption{The architecture of DUAL. (a) The embedding propagation layer aims to learn the node embeddings through message passing between adjacent nodes. (b) Each top-level module absorbs information from shared modules and its task-specific module in the embedding fusion layer. (c) The multi-task joint training layer aims to calculate the loss and end-to-end optimization for DUAL. (d) For the contrastive task, data-data pairs are derived from the embeddings of predictive ($\mathbf{Z}^{pre}$) and recommendation tasks ($\mathbf{Z}^{rec}$), respectively.}
\label{framework}
\end{figure*}

\section{The Proposed Model}
In this section, we propose DUAL, a novel self-supervised dual-auxiliary learning method. The framework is shown in Fig.\ref{framework}, which has three critical parts. 1) The embedding propagation layer aggregates meta-path-based neighbors by a GNN encoder to get the semantic-specific user/item embeddings. These embeddings will be selectively fused into final node representations for the recommendation and path-regression tasks. 2) The embedding fusion layer tells the difference of meta-paths through a gating network and obtains the optimally weighted combination of the semantic-specific node embedding for the recommendation and path-regression tasks. 3) The multi-task joint training layer simultaneously optimizes the recommendation and predictive (i.e., the path-regression) task, and then we maximize the agreement between node embeddings of them by a contrastive loss. 

Our model is trained on three tasks, namely recommendation task, predictive task, and contrastive task. As for recommendation task, we aim to estimate the likelihoods that users will interact with items. In addition to the recommended task, our model also combines a predictive task and a contrastive task to assist model training. The predictive task predicts the link-score, i.e., the normalized number of user-item paths induced by meta-paths, while the contrastive task maximizes the agreement between node representation for recommendation and predictive tasks. To avoid the negative transfer in multi-task joint training, we adopt CGC \cite{HongyanTang} as the bottom architecture of our model.

\subsection{Embedding Propagation Layer}
The input of the embedding propagation layer consists of shared modules (blue arrows) and task-specific modules (red arrows). Specifically, shared modules in DUAL are responsible for learning shared patterns, while task-specific modules extract patterns for specific tasks. As shown in Fig.\ref{framework}, shared modules (or task-specific modules) consist of multiple bipartite graphs and shared embedding $\mathbf{E}_{S}$ (or task-specific embedding $\mathbf{E}_{T}$). The number of bipartite graphs in modules is a hyper-parameter to tune. We use two bipartite graphs as input, one constructed through the interaction matrix ($\mathcal{G}_{UI}$) and the other constructed through a meta-path $\Phi$ starting from users and ending with items ($\mathcal{G}_{\Phi}$). The node embeddings after propagation of modules $(\mathbf{Z}_{(S,UI)}, \mathbf{Z}_{(S,\Phi)}, \mathbf{Z}_{(T,UI)}, \mathbf{Z}_{(T,\Phi)})$ are shown as follows:
\begin{equation}
\mathbf{Z}_{(S, U I)}=H\left(\mathbf{E}_{S}, \mathcal{G}_{U I}\right), \quad \mathbf{Z}_{(S, \Phi)}=H\left(\mathbf{E}_{S}, \mathcal{G}_{\Phi}\right),
\end{equation}
\begin{equation}
\mathbf{Z}_{(T, U I)}=H\left(\mathbf{E}_{T}, \mathcal{G}_{U I}\right), \quad \mathbf{Z}_{(T, \Phi)}=H\left(\mathbf{E}_{T}, \mathcal{G}_{\Phi}\right),
\end{equation}
where $H(\cdot)$ is the encoder function to encode connectivity information into representation learning. The encoder aims to perform convolutional operations on the input graph data, capturing fine-grained node preferences through message passing between adjacent nodes. Since LightGCN \cite{LightGCN} has superior performance for recommender systems, we adopt it as the basic structure of our encoder. It is worth noting that LightGCN adopts a layer combination operation to replace the self-loop connection of nodes. However, to avoid the negative effects of outliers in message passing, we remove the layer combination and maintain self-loops. The message passing operation (two propagation layers) with self-loops is defined as follows: 
\begin{gather}
\mathbf{Z}=\left(\hat{D}^{-\frac{1}{2}} \hat{A} \hat{D}^{-\frac{1}{2}}\right)\left(\left(\hat{D}^{-\frac{1}{2}} \hat{A} \hat{D}^{-\frac{1}{2}}\right) \mathbf{E} \right),\notag \\
\hat{D}=D+I,\notag \\
\hat{A}=A+I,
\end{gather}
where $I$ is the identity matrix extracted from bipartite graph $\mathcal{G}$. We denote $D$ and $A$ as the diagonal node degree matrix and the adjacency matrix, respectively. $\mathbf{E}$ is an initialized embedding matrix ($\mathbf{E}_{S}$ or $\mathbf{E}_{T}$).  

\subsection{Embedding Fusion Layer}
In the embedding fusion layer, each top-level module absorbs knowledge from shared modules and its task-specific module and fuses them, which indicates that the parameters of shared modules are influenced by both recommendation and path-regression predictive tasks. In contrast, parameters of task-specific modules are only affected by the corresponding specific task $k\in\{rec,pre\}$. $rec$ and $pre$ are abbreviations for recommendation and path-regression predictive tasks, respectively.

The outputs of shared modules and task $k$’s own module are combined through a gate structure to obtain fusing weight for the specific task. The structure of the gating network is based on a single-layer feedforward network with softmax as the activation function. More precisely, the output of task $k$’s gating network is formulated as:
\begin{equation}
\mathbf{Z}^{k}=g^{k}(x) S^{k},
\end{equation}
where $\mathbf{Z}^{k}$ is the final representations of all nodes of task $k$. $g^{k}(x)$ is a weighting function to calculate the weight vector of task $k$ through linear transformation and a softmax layer:
\begin{equation}
\label{weight}
g^{k}(x)=softmax\left(W^{k} x\right),
\end{equation}
where $x \in \mathbb{R}^{d}$ is the input representation. $W^{k} \in \mathbb{R}^{\left(m_{s}+1\right) \times d}$ is a parameter matrix, and $(m_{s}+1)$ is the number of shared modules and task $k$’s own module. $S^{k}$ is a selected matrix composed of all selected vectors including task $k$’s specific module ($\mathbf{Z}_{(T,UI)}$ or $\mathbf{Z}_{(T,\Phi)}$) and shared modules ($\mathbf{Z}_{(S,UI)}$ and $\mathbf{Z}_{(S,\Phi)}$):
\begin{equation}
\label{con_emb1}
S^{rec}= [\mathbf{Z}^{\top}_{(T,UI)}, \mathbf{Z}^{\top}_{(S,UI)}, \mathbf{Z}^{\top}_{(S,\Phi)}]^{\top},
\end{equation}
\begin{equation}
\label{con_emb2}
S^{pre}= [\mathbf{Z}^{\top}_{(T,\Phi)}, \mathbf{Z}^{\top}_{(S,UI)}, \mathbf{Z}^{\top}_{(S,\Phi)}]^{\top}.
\end{equation}

\subsection{Multi-Task Joint Training Layer}
The multi-task joint training layer first generates two scores indicating the user's preference for a item and the number for path instances between them. For each task $k$, a specific output layer is employed. We denote the representations of user $u$ and item $v$ in $\mathbf{Z}^{k}$ as $\mathbf{p}_{u}^{k}\in \mathbb{R}^{d}$ and $\mathbf{q}_{v}^{k}\in \mathbb{R}^{d}$. The scores of user $u$ for item $v$ are calculated as follows:
\begin{equation}
\label{R_hat}
\hat{R}_{u v}=\mathbf{h}^{rec\top}(\mathbf{p}_{u}^{rec} \odot \mathbf{q}_{v}^{rec})=\sum_{i=1}^{d} h_{i}^{rec} p_{u, i}^{rec} q_{v, i}^{rec},
\end{equation}
\begin{equation}
\label{M_hat}
\hat{M}_{u v}=\mathbf{h}^{pre\top}(\mathbf{p}_{u}^{pre} \odot \mathbf{q}_{v}^{pre})=\sum_{i=1}^{d} h_{i}^{pre} p_{u, i}^{pre} q_{v, i}^{pre},
\end{equation}
where $\mathbf{h}^{rec}\in \mathbb{R}^{d}$ and $\mathbf{h}^{pre}\in \mathbb{R}^{d}$ denote the output layers for recommendation and the predictive task, respectively. We construct a recommendation list of candidate items by descending $\hat{R}_{u v}$ for the recommendation task. 

\subsubsection{Loss Function of Recommendation}
We apply the efficient non-sampling learning \cite{Non-sampling} to optimize our target task – recommendation. It is a state-of-the-art learning method that uses only positive samples to guide model training without negative sampling. It has been confirmed in other studies \cite{chen2021graph, Non-sampling} that it has better performance and efficiency than traditional sampling learning methods. The non-sampling algorithm treats all node pairs that have not interacted as negative samples and incorporates them into the model's training. Specifically, the recommendation task loss is
\begin{equation}
\label{loss_rec}
\begin{aligned}
&\mathcal{L}_{rec}=\sum_{u \in \mathcal{B}} \sum_{v \in \mathcal{I}^{+}}\left(\left(c_{v}^{+}-c_{v}^{-}\right) \hat{R}_{u v}^{2}-2 c_{v}^{+} \hat{R}_{u v}\right)+\\
&\sum_{i=1}^{d} \sum_{j=1}^{d}\left(\left(\sum_{u \in \mathcal{B}} p_{u, i}^{rec} p_{u, j}^{rec}\right)\left(\sum_{v \in \mathcal{I}} c_{v}^{-} q_{v, i}^{rec} q_{v, j}^{rec}\right)\left(h_{i}^{rec} h_{j}^{rec}\right)\right).
\end{aligned}
\end{equation}
where we denote $\mathcal{B}$ as a batch of users and $\mathcal{I}$ as the set of all items. In the derivation of the non-sampling learning, each sample pair $(u, v)$ is given a weight $c_{uv}$. To facilitate the calculation, we follow the setting in the previous work \cite{Non-sampling} and set the weight of all positive and negative sample pairs to $c_{v}^{+}$ and $c_{v}^{-}$, respectively. They are two hyper-parameters. $\mathcal{I}^{+}$ is the positive samples set.

\subsubsection{Loss Function of Predictive Task}
Figure 1 shows a strong positive correlation between the interaction and the number of user-item paths induced by meta-paths. The higher the number of path instances between a user and an item, the higher the likelihood of an interaction between them. Predicting the number of user-item connections is beneficial for learning node preference. As a result, we model the number of user-item paths through a regression task.

As for the path-regression predictive task, we use the link-score matrix $\boldsymbol{M}$ as explicit supervisory signals. We minimize the mean squared error (MSE) between the predictions $\boldsymbol{\hat{M}}$ and the link-score matrix $\boldsymbol{M}$:
\begin{equation}
\label{mse}
\begin{aligned}
\mathcal{L}_{pre}=\sum_{u \in \mathcal{B}} \sum_{v \in \mathcal{I}} \left(M_{u v}-\hat{M}_{u v}\right)^{2}.
\end{aligned}
\end{equation}

Using the link-scores to expand supervision can provide more signals to guide the embedded learning than using only history interaction. The predictive task enhances the effectiveness of recommendations by fitting the intimate relationship between nodes.

\subsubsection{Loss Function of Contrastive Task}

Influenced by supervised signals, the trained $\mathbf{Z}^{rec}$ contains the node preference for historical interaction. In contrast, $\mathbf{Z}^{pre}$, another perspective of node relation, reflects the intimacy degree between nodes following a specific meta-path. Maximizing mutual information allows the intimate relationship between nodes to be further integrated into the $\mathbf{Z}^{rec}$ for the recommendation task. In addition, through tailored sampling methods, users can be as close as possible to items which strongly semantically relevant but have not seen before. In this way, the model makes recommendations based on semantically relevant items, not just historical ones.

As for the path-guided contrastive task, we maximize the agreement between node representation in $\mathbf{Z}^{rec}$ and $\mathbf{Z}^{pre}$. We treat these node representations as relevant but distinct views. Then, we employ a contrastive objective that distinguishes the embeddings of positive and negative samples in the two views. Specifically, as shown in Fig.\ref{framework} any node in $\mathbf{Z}^{rec}$ is treated as an anchor ($v_{1}$). In $\mathbf{Z}^{pre}$, the same anchor node ($v'_{1}$) and nodes whose link-score between themselves and the anchor is 1 ($v'_{2}, v'_{3}$), are positive samples ($v'_{1}, v'_{2}, v'_{3}$). Nodes whose link-score between themselves and the anchor is 0 are naturally regarded as negative samples ($v'_{4}, v'_{6}$). Because these nodes are not closely related to the anchor, they are more probably to be truly negative instances. Our model selects positive and negative samples through the guidance of link-score, which can increase the number of positive samples and improve the credibility of negative samples.

For pairs of positive samples, we desire to maximize their consistency under different views. On the contrary, for negative sample pairs, it is necessary to enlarge the divergence between them, forcing them to be farther apart in the latent space. Specifically, we adopt the same form of loss function as SGL \cite{SGL} - contrastive loss to optimize the path-guided contrastive task:
\begin{equation}
\label{loss_con}
\begin{aligned}
\mathcal{L}_{con}=\sum_{(i, j) \in Q^{+}}-\log \frac{\exp \left(s \left(\mathbf{z}_{i}^{r e c}, \mathbf{z}_{j}^{p r e}\right) / \tau\right)}{\sum_{(i,\tilde{j}) \in Q^{-}} \exp \left(s \left(\mathbf{z}_{i}^{r e c}, \mathbf{z}_{\tilde{j}}^{p r e}\right) / \tau\right)}.
\end{aligned}
\end{equation}

We denote $Q^{+}$ as the set of positive samples and $Q^{-}$ as negative samples. The function $s(\cdot,\cdot)$ is designed to measure the distance between two node embeddings in the latent space. Concretely, we employ cosine similarity to calculate the similarity between nodes in a node pair. $\tau$ is the temperature hyperparameter.

\begin{algorithm}[t]
	\caption{The overall process of DUAL}
	\LinesNumbered 
	\label{alg:algorithm}
	\KwIn{User-item bipartite graphs $\mathcal{G}_{UI}$ and $\mathcal{G}_{\Phi}$ \qquad The weight of predictive task $\lambda_{pre}$ \qquad \qquad \qquad The weight of contrastive task $\lambda_{con}$ }
	\KwOut{Final node embeddings $\mathbf{Z}^{rec}$ and $\mathbf{Z}^{pre}$}
	\DontPrintSemicolon
	Randomly initialize node embedding $\mathbf{E}_{S}$ and $\mathbf{E}_{T}$.\\
	\While{not converge}{
	    \For{each epoch}{
    		\For{path = UI, $\Phi$}{
        		Calculate latent features $\mathbf{Z}_{(S,path)} \gets H(\mathbf{E}_{S}, \mathcal{G}_{path})$\\
        		Calculate latent features $\mathbf{Z}_{(T,path)} \gets H(\mathbf{E}_{T}, \mathcal{G}_{path})$\\
            }
            Concatenate the learned embeddings $S^{rec} \gets [\mathbf{Z}^{\top}_{(T,UI)}, \mathbf{Z}^{\top}_{(S,UI)}, \mathbf{Z}^{\top}_{(S,\Phi)}]^{\top}$ \\
            Concatenate the learned embeddings $S^{pre} \gets [\mathbf{Z}^{\top}_{(T,\Phi)}, \mathbf{Z}^{\top}_{(S,UI)}, \mathbf{Z}^{\top}_{(S,\Phi)}]^{\top}$ \\
            \For{k = rec, pre}{
        		Calculate the weight of meta-paths $g^{k} \gets softmax\left(W^{k} x\right)$\\
        		Fuse the final embedding $\mathbf{Z}^{k} \gets g^{k}(x) S^{k}$
            }
            Predict the probability $\hat{R}_{u v}$ according to Eq.(\ref{R_hat})\\
            Predict the link-score $\hat{M}_{u v}$ according to Eq.(\ref{M_hat})\\
    		Evaluate $\mathcal{L}_{con}$, $\mathcal{L}_{rec}$, and $\mathcal{L}_{pre}$ according to Eq.(\ref{loss_rec}) - Eq.(\ref{loss_con})\\
    		Jointly optimize the overall objective $\mathcal{L}  \gets \mathcal{L}_{rec} + \lambda_{pre} \mathcal{L}_{pre} + \lambda_{con} \mathcal{L}_{con}$ \\
    		Back propagation and update parameters\\
	    }
	}
    \textbf{return} $\mathbf{Z}^{rec}, \mathbf{Z}^{pre}$\\
\end{algorithm}

Throughout the whole model, we are most concerned about the performance of the recommendation, and auxiliary tasks are designed to make the recommendation profit. Therefore, we can further define the loss function of the whole model as the weighted summation of the recommendation's loss and auxiliary tasks' loss as follows:
\begin{equation}
\label{loss}
\mathcal{L}=\mathcal{L}_{rec} + \lambda_{pre} \mathcal{L}_{pre} + \lambda_{con} \mathcal{L}_{con},
\end{equation}
where $\lambda_{pre}$ denotes the weight of the predictive task, and $\lambda_{con}$ denotes the weight of the contrastive task. Both $\lambda_{pre}$ and $\lambda_{con}$ are hyper-parameters. The overall process of DUAL is presented in Algorithm \ref{alg:algorithm}.

Our framework contains no parameters on any model network except for gating networks and output layers. No parameters are needed to be trained in lightGCN, and the only trainable model parameters in the whole embedding propagation layer are node embedding matrices (shared node embedding $\mathbf{E}_{S}$ and task-specific node embedding $\mathbf{E}_{T}$ ). Therefore, our framework retains the low computational complexity.

\section{Experiments}

In this section, we first display the superiority of our DUAL model by extensive experiments on three public datasets and comparing state-of-the-art recommendation models. Second, we also perform ablation studies on the two auxiliary tasks to demonstrate the efficacy of the hybrid method. Besides, we explore the influence of different meta-paths and encoders on experimental results. Finally, we perform parameter sensitivity experiments to guide the choice of hyper-parameter for our DUAL model.

\begin{table}[htbp]
\caption{Statistics of the datasets.}
\footnotesize
\label{datasets}
\centering
\begin{tabular}{lllll}
\hline
\multirow{2}{*}{\begin{tabular}[c]{@{}l@{}}Datasets\\ (Density)\end{tabular}} &
\multirow{2}{*}{\begin{tabular}[c]{@{}l@{}}Relations \\ (A-B)\end{tabular}} &
\multirow{2}{*}{\begin{tabular}[c]{@{}l@{}}Number \\ of (A-B)\end{tabular}} &
\multirow{2}{*}{\begin{tabular}[c]{@{}l@{}}Number \\ of A\end{tabular}} &
\multirow{2}{*}{\begin{tabular}[c]{@{}l@{}}Number \\ of B\end{tabular}} \\
&                                 &                         &                         &                          \\ \hline
\multirow{3}{*}{\begin{tabular}[c]{@{}l@{}}Last-FM\\ (1.36\%)\end{tabular}}     & User-User  & 18,802       & 1,892  & 1,892  \\ 
                            & User-Artist & 92,834     & 1,892  & 17,632  \\ 
                             & Artist-Artist & 153,399   & 17,632 & 17,632 \\ \hline 
\multirow{3}{*}{\begin{tabular}[c]{@{}l@{}}Yelp\\ (0.54\%)\end{tabular}} & User-User & 158,590       & 16,239 & 16,239 \\ 
                             & User-Business & 198,397       & 16,239 & 14,284 \\ 
                             & Business-Category & 40,009      & 14,284 & 511  \\ \hline 
\multirow{3}{*}{\begin{tabular}[c]{@{}l@{}}Douban-Book\\ (0.33\%)\end{tabular}} & User-User & 169,150       & 12,748 & 12,748 \\ 
                             & User-Book & 792,026       & 13,024 & 22,347 \\ 
                             & Book-Author & 21,905     & 21,907 & 10,805  \\  \hline
\end{tabular}
\end{table}

\subsection{Experimental Setup}
\subsubsection{Datasets}
We experiment with three publicly accessible datasets from different domains: Last-FM\footnote{\label{note1}https://github.com/librahu/HIN-Datasets-for-Recommendation-and-Network-Embedding} dataset from music domain, Yelp$^{1}$ dataset from business domain, and Douban-book$^{1}$ dataset from book domain.
These datasets are frequently used to validate the effectiveness of previous recommendation models \cite{Hwang, YuJ}. We convert ratings in datasets into binary feedback of 0 and 1, which only distinguish whether the user and the item have interacted. Following previous works \cite{S3-rec}, we only keep active users and popular items with no less than five interaction records. In addition to the different recommendation scenarios, the datasets also have various data sparsity: The Douban-book and Yelp are very sparse datasets with a density of only 0.33\% and 0.54\%, respectively, while the density of Last-FM is 1.36\%, which is denser than that of Yelp and Douban-book. The statistics of these datasets are showed in Table \ref{datasets}. 


\subsubsection{Baselines}
To verify the superiority of our DUAL model, we compare it with some state-of-the-art models. It is worth noting that BUIR is a predictive model and SGL is a contrastive model.

\begin{itemize}
\item NeuMF \cite{NeuMF}: A classic recommendation framework that replaces the embedding inner product through a neural network. Each feature extraction module holds its independent representation.

\item SLIMElastic \cite{XNing}: A classic method in the top-N recommendation. It improves recommendation accuracy through an auxiliary sparse coefficient matrix learned during model training.
\item NGCF \cite{NGCF}: A graph-based collaborative filtering method. It can aggregate node embeddings in a graph structure constructed from interactive data, explicitly putting collaborative filtering signals into the embedding propagation process.
\item LINE \cite{LINE}: A network embedding model. It can map all nodes to a low-dimensional embedding while maintaining the original network structure as much as possible. The method can handle large-scale and various complex data.
\item DGCF \cite{DGCF}: A graph-based collaborative filtering method, which can mine and refine various intentions of users and learn the critical factors of interaction between the user and an item. 
\item LightGCN \cite{LightGCN}: An excellent graph-based method that removes damage recommendation's complex operation during graph convolution. It achieves double improvement in capability and efficiency of recommendation by simplifying NGCF.
\item BUIR \cite{BUIR}: A framework for collaborative filtering. It only uses positive samples to guide model training and alleviates the problem of data sparsity through random data augmentation. This method is a predictive self-supervised learning model.
\item SGL \cite{SGL}: A framework based on SSL, which performs data augmentation on the input bipartite graph with a dropout of nodes and edges. Then it draws on the idea of contrastive learning to construct an SSL task.
\end{itemize}

\subsubsection{Evaluation Metrics}
In order to more intuitively demonstrate the superiority of the DUAL model, we use two typical evaluation metrics, Recall@K and NDCG@K, to evaluate the model's performance. These two evaluation metrics are widely applied in recommendation methods \cite{LightGCN, NGCF, DGCF}. Recall focuses on the coverage of positive samples in the recommendation list, and NDCG evaluates sorting accuracy.

\begin{table*}[htbp]
\caption{Performance comparison of DUAL model and baselines on three public datasets. Bold numbers represent optimal performance, and underlined numbers indicate suboptimal results.}
\label{comparison}
\centering
\scalebox{1}{
\begin{tabular}{llllllllll}
\hline
Dataset & Method    & Recall@5 & Recall@10 & Recall@15 & Recall@20 & NDCG@5  & NDCG@10 & NDCG@15 & NDCG@20 \\ \hline
\multirow{7}{*}{Last-FM}     
        & NeuMF & 0.1782 & 0.2561 & 0.3145 & 0.3633 & 0.1804 & 0.2175 & 0.2403 & 0.2570 \\ 
        & SLIMElastic & 0.1370 & 0.1952 & 0.2328 & 0.2660 & 0.1442 & 0.1721 & 0.1871 & 0.1986 \\ 
        & NGCF      & 0.1658  & 0.2456   & 0.3051   & 0.3516   & 0.1643 & 0.2021 & 0.2254 & 0.2415 \\ 
        & LINE      & 0.1703  & 0.2463   & 0.2964   & 0.3422   & 0.1709 & 0.2069 & 0.2265 & 0.2422 \\ 
        & DGCF      & 0.1789  & 0.2610   & 0.3252    & 0.3702   & 0.1824 & 0.2216  & 0.2461 & 0.2616  \\ 
        & LightGCN  & 0.1853   & {\ul 0.2738}   & {\ul 0.3290}   & 0.3745   & 0.1885 & 0.2299 & 0.2516 & 0.2675 \\ 
        & BUIR      & {\ul 0.1947}  & 0.2735   & 0.3277   & 0.3740   & 0.1985 & 0.2352 & 0.2561 & 0.2719 \\ 
        & SGL      & 0.1845  & 0.2695   & 0.3287   & {\ul 0.3785}   & {\ul 0.2128} & {\ul 0.2456} & {\ul 0.2701} & {\ul0.2880} \\ \cline{2-10}
        & \textbf{DUAL} & \textbf{0.2052}  & \textbf{0.2864}    & \textbf{0.3436}   & \textbf{0.3930}    & \textbf{0.2250} & \textbf{0.2533} & \textbf{0.2767} & \textbf{0.2941} \\ 
        & Improv.         & 5.39\%  & 4.60\%   & 4.44\%   & 3.83\%   & 5.73\% & 3.14\% & 2.44\%  & 2.12\% \\ \hline
\multirow{7}{*}{Yelp} 
        & NeuMF & 0.0366 & 0.0582 & 0.0794 & 0.0980 & 0.0361 & 0.0423 & 0.0491 & 0.0547 \\ 
        & SLIMElastic & 0.0253 & 0.0388 & 0.0500 & 0.0595 & 0.0248 & 0.0286 & 0.0320 & 0.0349 \\ 
        & NGCF      & 0.0331 & 0.0528 & 0.0732 & 0.0905 & 0.0313 & 0.0376 & 0.0439 & 0.0490  \\ 
        & LINE      & 0.0367 & 0.0545 & 0.0713 & 0.0894 & 0.0335 & 0.0388 & 0.0444 & 0.0496 \\ 
        & DGCF      & 0.0424 & 0.0690 & 0.0885 & 0.1089 & 0.0394 & 0.0478 & 0.0539 & 0.0598  \\ 
        & LightGCN  & 0.0430 & {\ul 0.0707} & {\ul 0.0936} & {\ul 0.1124} & 0.0426 & 0.0508 & 0.0580 & 0.0635 \\ 
        & BUIR      & 0.0358 & 0.0583 & 0.0782 & 0.0965 & 0.0374 & 0.0450 & 0.0520 & 0.0580  \\ 
        & SGL      & {\ul 0.0441}  & 0.0700   & 0.0917   & 0.1103   & {\ul 0.0456} & {\ul 0.0537} & {\ul 0.0609} & {\ul 0.0671} \\ \cline{2-10}
        & \textbf{DUAL} & \textbf{0.0461}  & \textbf{0.0776}    & \textbf{0.1033}   & \textbf{0.1258}    & \textbf{0.0480} & \textbf{0.0574} & \textbf{0.0656} & \textbf{0.0724} \\ 
        & Improv.         & 4.44\% & 9.73\%   & 10.32\%   & 11.90\%   & 5.26\% & 6.85\% & 7.68\%  & 7.93 \%\\ \hline
\multirow{7}{*}{Douban-book} 
        & NeuMF & 0.0615 & 0.0929 & 0.1175 & 0.1404 & 0.0764 & 0.0826 & 0.0893 & 0.0962 \\ 
        & SLIMElastic & 0.0787 & 0.1123 & 0.1326 & 0.1494 & 0.1141 & 0.1141 & 0.1174 & 0.1215 \\ 
        & NGCF      & 0.0709  & 0.1053   & 0.1299    & 0.1495   & 0.0850 & 0.0920  & 0.0990 & 0.1051  \\ 
        & LINE      & 0.0753  & 0.1050   & 0.1235   & 0.1389   & 0.1009 & 0.1046 & 0.109 & 0.1136 \\ 
        & DGCF      & 0.0842  & 0.1230   & 0.1506    & 0.1734   & 0.1089 & 0.1149  & 0.1218 & 0.1287  \\ 
        & LightGCN  & 0.0917  & 0.1344   & 0.1624   & 0.1870    & 0.1197 & 0.1263 & 0.1332 & 0.1404 \\ 
        & BUIR      & 0.0622  & 0.0963   & 0.1235   & 0.1445   & 0.0853 & 0.0899 & 0.0968 & 0.1029 \\ 
        & SGL      & {\ul 0.0952}  & {\ul 0.1372}   & {\ul 0.1696}   & {\ul 0.1960}   & {\ul 0.1414} & {\ul 0.1445} & {\ul 0.1520} & {\ul 0.1599} \\ \cline{2-10}
        & \textbf{DUAL} & \textbf{0.1072}  & \textbf{0.1520}    & \textbf{0.1827}   & \textbf{0.2078}    & \textbf{0.1641} & \textbf{0.1631} & \textbf{0.1683} & \textbf{0.1746} \\ 
        & Improv.         & 12.60\% & 10.78\%   & 7.72\%   & 6.02\%   & 16.05\% & 12.84\% & 10.72\%  & 9.19 \%\\ \hline
\end{tabular}}
\end{table*}

\subsubsection{Implementation Details}
We randomly divide each public dataset and use 80\% of the data for model training, 10\% as the valid set, and the rest as the test set. The valid and test data are respectively used to determine model parameters and evaluate the final model. The parameter settings of baselines are the same as those in their paper and then tuned for optimal performance. For the fairness of the comparison, we set the embedding dimension of all models to 256. Additionally, we set the patience of the early stopping method to 20, i.e., models stop training if the performance does not improve for 20 consecutive epochs. For models employing the sampling strategy, we uniformly set the number of negative samples to 5.

For the DUAL model, we randomly initialize parameters and optimize the model with Adam \cite{Adam}. We set the learning rate to 0.001, and the dropout ratio was set to 0.3 to prevent overfitting. The weight of non-zero instances $c_{v}^{+}$ is set to 1 for all datasets. The negative weight $c_{v}^{-}$ is set to 0.15, 0.2, and 0.25 for Last-FM, Yelp, and Douban-book, respectively. We set the weight of the path-regression task $\lambda_{pre}$ as 0.3 for Last-FM, 0.03 for Yelp and Douban-book. Due to the different loss functions of path-regression and path-guided contrastive tasks, the weight of the path-guided contrastive task $\lambda_{con}$ is much greater than that of the path-regression task. Therefore, we give a lower weight to the contrastive task to balance the training rate between different tasks. Specifically, the weight of the contrastive task $\lambda_{con}$ is set to 5e-4 for Last-FM, 1e-7 for Yelp, and 1e-5 for Douban-book. For Last-FM, Yelp, and Douban-book datasets, we select the optimal meta-path in [UUA, UAA, UAUA], [UUB, UBCB], and [UBAB, UBUB], respectively.

\subsection{Performance Comparison}
The performance results of DUAL and baselines are listed in Table \ref{comparison}. In experiments, we set $K$ of Racall@K and NDCG@K as 5, 10, 15, and 20 to evaluate the DUAL and baselines. According to the results, the following results can be analyzed: 

First, the DUAL outperforms all baselines on three public datasets, achieving the best performance. The average improvement of our DUAL model over the best baseline is 3.96\% on the Last-FM dataset, 8.02\% on the Yelp dataset, and 10.74\% on the Douban-book dataset, which validates the superiority of the DUAL model. The improvements on Yelp and Douban-book are more significant than those on the Last-FM dataset, which we hold due to the datasets' different densities. Yelp and Douban-book datasets are more sparse and lack of interaction data for model training compared with the Last-FM dataset. 

We believe that the superior performance of DUAL comes from the following three aspects: 1) We adopt a multi-task sharing structure, which can explicitly separate shared modules and task-specific modules, extract and separate semantic knowledge reflected by social relationships and item categories, through multi-tasking Joint representation learning improves model performance. 2) DUAL can accurately capture the strong correlation between the number of path instances following the specific meta-path and user-item interactions. 3) Combining different types of SSL tasks can provide more semantic supervisory signals for recommendations, which is essential in improving the model's performance.

Second, methods based on LightGCN generally outperform other graph-based. For example, in Table \ref{comparison}, LightGCN, SGL, and DUAL outperform NGCF and DGCF. This conclusion is consistent with previous work \cite{LightGCN}, which indicates the strong performance of LightGCN in the recommendation. LightGCN has such strong performance because it is the progressive graph-based CF model, which designs a light graph convolution to ease the training difficulty and pursue better generative ability.

Third, self-supervised learning methods (BUIR, SGL, and DUAL) generally perform better than supervised methods. It can be seen that our hybrid approach that jointly applies contrastive task and predictive task significantly outperforms the two models that only use one standalone predictive task (BUIR) or contrastive task (SGL). It verifies the effectiveness of both types of auxiliary tasks in modeling user preferences. In addition, the self-supervised tasks of BUIR and SGL models, which are universal self-supervised tasks, originate from computer vision and graph learning, respectively. Our model DUAL outperforms these two baselines, proving that self-supervised task closely related to recommendation is superior to general self-supervised tasks. Regarding the poor performance of the self-supervised model BUIR on Yelp and Douban-book datasets, we speculate that because BUIR is only trained with positive samples, it does not work well on extremely sparse datasets.

\subsection{Ablation Study}

In this section, we conduct ablation studies on the path-regression task and path-guided contrastive task to demonstrate the efficacy of our DUAL model. Since the contrastive task is based on the predictive task, the contrastive task is also removed in the variant that eliminates the predictive task. 
\begin{itemize}
\item DUAL-PC: The variant model of DUAL without predictive and contrastive tasks.
\item DUAL-C: The variant model of DUAL without contrastive task.
\end{itemize}

\begin{table}[htbp]
\caption{Performance of variants of DUAL on Last-FM dataset.}
\label{variants}
\centering
\scalebox{1.1}{
\begin{tabular}{lllll}
\hline
Model & recall@10  & recall@20 & ndcg@10 & ndcg@20 \\ \hline
DUAL-PC & 0.2823       & 0.3875       & 0.2477     & 0.2896     \\
DUAL-C & 0.2845        & 0.3906        & 0.2498     & 0.2910      \\
DUAL   & \textbf{0.2864} & \textbf{0.3930}    & \textbf{0.2533}     & \textbf{0.2941}  \\ \hline
\end{tabular}}
\end{table}

\begin{figure*}[htbp]
\centerline{\includegraphics[scale=0.4]{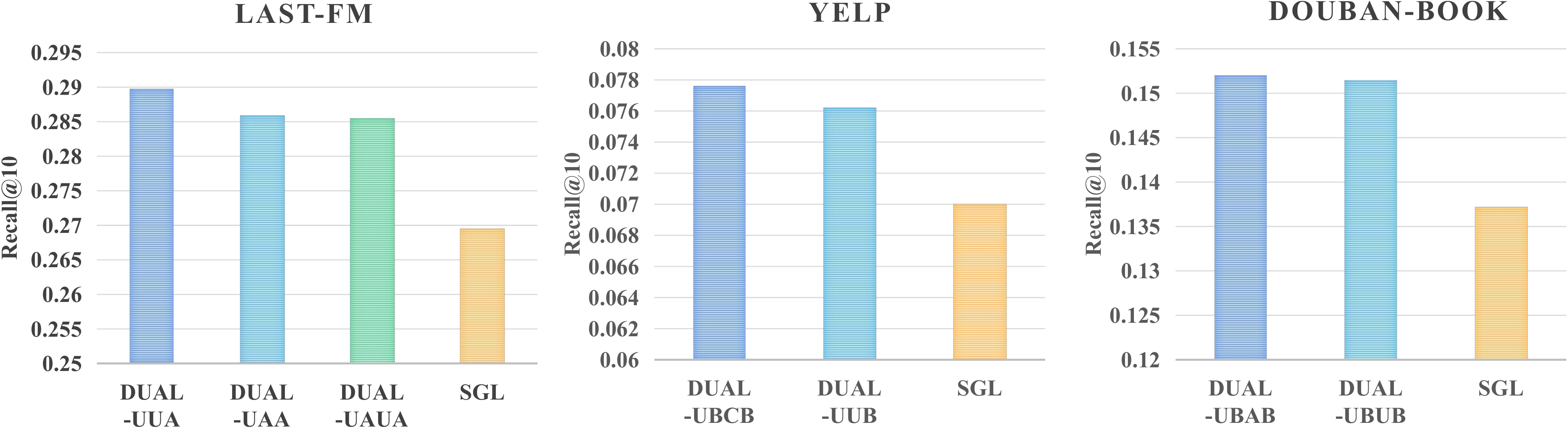}}
\caption{Impacts of different meta-paths. We abbreviate user and artist for the Last-FM to U and A, respectively. For the Yelp, we abbreviate user, businesses, and category to U, B, and C, respectively. For the Douban-book, we abbreviate user, book, and author to U, B, and A, respectively.}
\label{metapath}
\end{figure*}

Table \ref{variants} displays the performance of different variants of DUAL. As shown in the table, the predictive task or the contrastive task leads to better recommendation results. The performance of our DUAL is further improved when all two SSL tasks are applied. It is verified that the combination of path-regression and edeg-guided contrastive task is more conducive to recommendation than a single type of SSL task.

The various tasks in real-world recommender systems are usually competitive and complex. Therefore, the negative transfer is common in multi-task learning, especially for loosely correlated tasks \cite{transfer}. However, our proposed model DUAL has a multi-task sharing structure that explicitly separates shared and task-specific components to bypass the negative transfer phenomenon. The experimental results verify that the DUAL model can simultaneously train multiple tasks in one framework and improve the recommendation effect through information sharing between tasks.

\subsection{Impacts of Different Meta-Paths}
To further evaluate our DUAL model, we conduct an experiment on impacts of using different meta-paths. As shown in Fig.\ref{metapath}, the DUAL with different meta-paths performs better than the best performing baseline SGL, which means combining meta-paths into the framework is effective for improving recommendations. The DUAL can accurately capture the strong correlation between user-item interactions and the number of connections following the specific meta-paths. Furthermore, different meta-paths have tiny different performance improvements. For example, for the Last-FM dataset, users have similar preferences with their friends connected by social relationships (UUA). For the Yelp dataset, users are more interested in businesses within the same category (UBCB), while for the Douban-book dataset, users are more inclined to select books by the same author as the books they have previously read (UBAB).

\subsection{Impacts of Different Encoders}
Architecturally, the proposed DUAL is model-agnostic to facilitate various graph-based recommendation methods. We evaluate the DUAL modle with five graph neural networks: GCN \cite{GCN}, GAT \cite{GAT}, TAGCN \cite{TAGCN}, SGC \cite{SGC}, and LightGCN \cite{LightGCN}. As we expected, the results in Table \ref{encoder} demonstrate that the effectiveness of the framework differs depending on graph encoders. However, it is worth noting that three variants of our model (DUAL-GCN, DUAL-SCG, and DUAL) have better performance than the best baseline SGL. It shows that our proposed model DUAL has excellent extensibility and versatility and can be easily transplanted to other graph neural networks to achieve excellent recommendation performance. In particular, the model effect based on LightGCN is outstanding. That is because LightGCN, as a graph neural network specially designed for the recommender system, removes the useless or harmful operations and is more suitable for the field of recommendation.

\begin{table}[htbp]
\caption{Performance of DUAL trained by various encoders on the Last-FM dataset.}
\label{encoder}
\centering
\scalebox{1}{
\begin{tabular}{lllll}
\hline
Variants      & recall@10 & recall@20 & ndcg@10 & ndcg@20 \\ \hline
DUAL-TAGCN       & 0.2523    & 0.3487    & 0.2178  & 0.2556  \\
DUAL-GAT         & 0.2520    & 0.3522    & 0.2190  & 0.2579  \\
DUAL-GCN         & 0.2831    & 0.3861    & 0.2485  & 0.2879  \\ 
DUAL-SGC         & 0.2806    & 0.3878    & 0.2489  & 0.2899  \\
DUAL (LightGCN)    & \textbf{0.2864}    & \textbf{0.3930}    & \textbf{0.2533}  & \textbf{0.2941}  \\ \hline
SGL         & 0.2695    & 0.3787    & 0.2456  & 0.2880  \\\hline
\end{tabular}}
\end{table}

\subsection{Hyper-Parameter Sensitivity}
To guide the parameter choice of the DUAL model, we perform hyper-parameter studies of the result of DUAL. Specifically, the dataset used to evaluate the model performance is Last-FM, and LightGCN is applied as the encoder of DUAL. We test the result changes of the DUAL model concerning hyper-parameters on the number of layers and link-score of negative samples for contrastive tasks.

\begin{figure}[htbp]
\centerline{\includegraphics[scale=0.37]{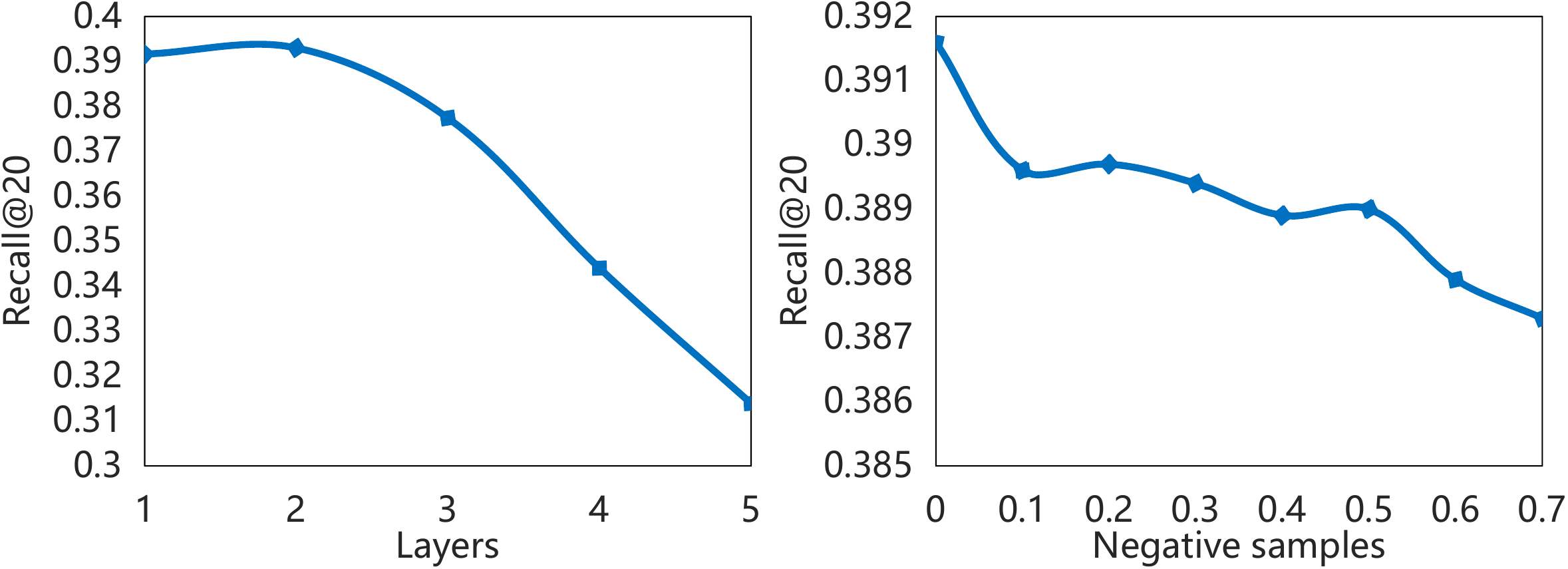}}
\caption{Performance of DUAL with regard to the number of layers and the link-score of negative samples on the Last-FM dataset.}
\label{parameter}
\end{figure}

For the number of convolutional layers of DUAL, we choose its range in the range of [1, 2, 3, 4, 5] and plot the results in Fig.\ref{parameter}. The result demonstrates that the effect of DUAL is relatively more susceptible to the number of layers than the link-score of negative samples. As can be seen from Fig.\ref{parameter}, the model performance is improved by increasing the number of convolutional layers from 1 to 2. Usually, two propagation layers are adequate to capture heterogeneous signals.DUAL suffers the over-smoothed problem when the layer number exceeds two.

For the link-score of negative samples for the contrastive task, we select it in the range of [0, 0.1,...,0.7]. As shown in Fig.\ref{parameter}, DUAL is susceptible to the link-score of negative samples, and we need to choose it carefully for the best result. Usually, small values of link-score can lead to ideal results. This is because a large link-score between a user and an item means that they are more closely related. Therefore, the probability that the item is a potential positive is higher. Selecting these samples as negative samples will introduce noise into the model training, which degrades the recommendation performance. Our model provides a new scheme for selecting negative samples, i.e., by adjusting the link-score of negative samples to avoid selecting potentially positive samples. The experimental results also verify that the framework is effective.

\section{Conclusion and Future Work}
In this work, we propose a self-supervised dual-auxiliary learning method (DUAL), which automatically provides the additional supervisory signals and fully integrates rich semantic information reflected by social relationships and item categories to improve the model's performance. In our approach, we extract the commuting matrix of a specific meta-path and normalize it to obtain the link-score. Through the link-score, the path-regression predictive and path-guided contrastive tasks can be naturally closely integrated. Additionally, our proposed model DUAL effectively combines two auxiliary tasks while avoiding negative transfer. Experimental results on public datasets demonstrate that DUAL exceeds state-of-the-art supervised and self-supervised recommendation methods.

We will study how to develop a simplified SSL framework with multiple meta-paths in the future. Besides, we will also consider applying the framework to more complicated recommendation scenarios, such as bundle or group recommendation.

\section*{Acknowledgment}
This work was supported by National Key R\&D Program of China (2018YFB1403602), the National Natural Science Foundation of China (62176028), and the Overseas Returnees Innovation and Entrepreneurship Support Program of Chongqing (cx2020097).

\printbibliography


 

\vspace{-33pt}

\begin{IEEEbiography}[{\includegraphics[width=1in,height=1.25in,clip,keepaspectratio]{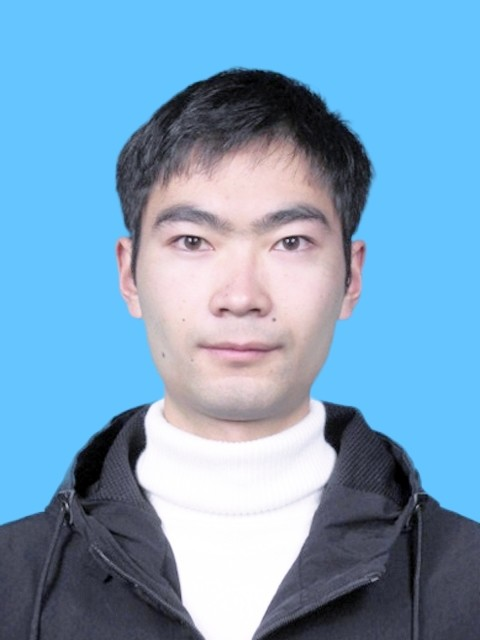}}]{Yinghui Tao}
received the B.S. degree from Guizhou University, Guizhou, China, in 2020, where he is currently pursuing the master’s degree with the School of Big Data \& Software Engineering, Chongqing University. His research interests include recommender systems and anomaly detection.
\end{IEEEbiography}

\begin{IEEEbiography}[{\includegraphics[width=1in,height=1.25in,clip,keepaspectratio]{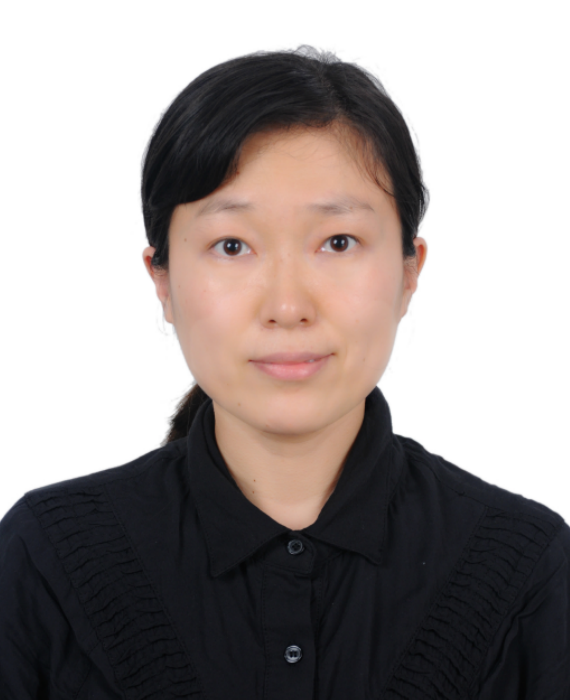}}]{Min Gao}
received the MS and PhD degrees in computer science from Chongqing University in 2005 and 2010 respectively. She is an associate professor at the School of Big Data \& Software Engineering, Chongqing University. She was a visiting researcher at University of Reading and Arizona State University. Her research interests include recommendation systems, service computing, and data mining. She has published over 40 refereed journal and conference papers in these areas. She has grants from the National Natural Science Foundation of China, the China Postdoctoral Science Foundation, and the China Fundamental Research Funds for the Central Universities. She is member of the IEEE and China Computer Federation (CCF).
\end{IEEEbiography}

\begin{IEEEbiography}[{\includegraphics[width=1in,height=1.25in,clip,keepaspectratio]{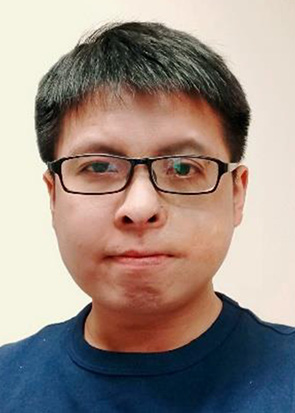}}]{Junliang Yu}
received the B.S. and MS degree in Software Engineering from Chongqing University, Chongqing, China, in 2013 and 2018, respectively. Currently, he is a Ph.D. student with the school of Information Technology and Electrical Engineering at the University of Queensland, Queensland, Australia. His research interests include recommender systems and anomaly detection.
\end{IEEEbiography}

\begin{IEEEbiography}[{\includegraphics[width=1in,height=1.25in,clip,keepaspectratio]{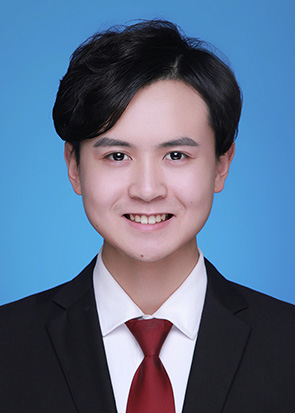}}]{Zongwei Wang}
received the B.S. and MS degree in Software Engineering from Chongqing University in 2018 and 2021 respectively. His research interests include recommender systems and anomaly detection.
\end{IEEEbiography}

\begin{IEEEbiography}[{\includegraphics[width=1in,height=1.25in,clip,keepaspectratio]{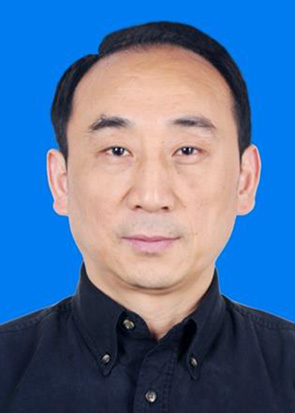}}]{Qingyu Xiong}
received the B.S. and M.S. degree from the School of Automation, Chongqing University, in 1986 and 1991, respectively, and the Ph.D. degree from the Kyushu University of Japan in 2002. He is currently the Dean of the School of Big Data \& Software Engineering, Chongqing University. His research interests include neural networks and their applications. He has authored over 100 journal and conference papers in these areas. He has over 20 research and applied grants.
\end{IEEEbiography}

\begin{IEEEbiography}[{\includegraphics[width=1in,height=1.25in,clip,keepaspectratio]{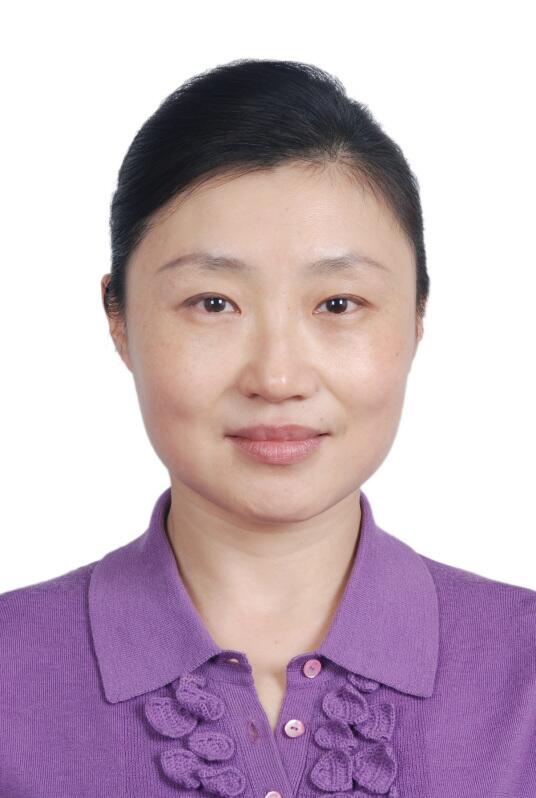}}]{Xu Wang}
received the M.S. and Ph.D. degrees in Mechanical Manufacturing from Chongqing University, Chongqing, China. She used to be the deputy secretary of the party committee of Chongqing University and is currently a professor at the College of Mechanical and Vehicle Engineering, Chongqing University. Her research interests include modern logistics and supply chain management, industrial Internet, service science and engineering.
\end{IEEEbiography}

\vspace{11pt}


\vfill

\end{document}